\documentstyle[preprint,aps]{revtex}
\draft
\begin{document}               
\title{
Generalized yrast states of a Bose condensate in harmonic trap
for a universality class of repulsive interactions
}
\author{M. S. Hussein and  O.K. Vorov
}
\address{
Instituto de Fisica, 
Universidade de Sao Paulo \\
Caixa Postal 66318,  05315-970,  \\
Sao Paulo, SP, Brasil
}
\date{27 February 2001}
\maketitle
\begin{abstract}
For a system of $N$ bosons in a 2d harmonic trap with frequency
$\omega$, interacting
via repulsive forces $V\ll\hbar\omega$, we develop supersymmetric
method
to find 
the lowest energy states
of rotating Bose condensate as function
of two quantum numbers, 
the total angular momentum and
the angular momentum of internal excitations
(generalized yrast states).
The energies
of these 
condensed vortex
states are expressed through 
the 
single two-body matrix element of interaction $V$. 
A broad {\it universality} class of the repulsive 
interactions for which these results 
hold
is described 
by a simple integral condition on $V$.
It includes Gaussian, $\delta$-function and $log$-Coulomb forces. 
\end{abstract}

Creation of traps for cold rarified atoms \cite{ANDERSON}, 
which has made possible experimental
investigation 
of Bose-Einstein condensation,
has 
stimulated
theoretical studies on the systems of weakly interacting 
bosons confined by 
parabolic potential
\cite{nature,Wilkin,Mottelson,BP,REVIEW}. 
Among the important questions about behavior of such systems
is their response to rotation and the onset of vorticity
\cite{nature,Wilkin,Mottelson,BP,REVIEW}.
In this context, of great importance is the structure 
and the spectrum of
the ground states of rotating condensate at given angular momentum,
the {\it yrast states}
\cite{nature,Wilkin,Mottelson,BP}.

The particular problem of wide recent 
interest\cite{Wilkin,Mottelson,BP}
arises
in the weak coupling limit, which
is hoped 
\cite{nature,Wilkin,Mottelson}
to be reached 
experimentally
in nearest future. 
One considers $N$ spinless bosons in spherical 2d harmonic 
trap\cite{3d2d}. 
The noninteracting system 
has equidistant spectrum 
$\hbar\omega n$
of high degeneracy\cite{Mottelson}
which grows exponentially with $n$ at $n-N\gg1$. This degeneracy
is related to the number of ways to distribute the total 
energy $\hbar\omega n$
among the 
particles.
The short-range interactions $V(r)$ 
between the 
atoms are assumed weak 
in the sense that 
hoppings
between different
$\hbar\omega n$
levels can be neglected,
and the problem still requires nonperturbative
solution for the highly degenerate states at single level 
$\hbar\omega n$, which is similar in spirit to the problem of
Landau level for the electrons in magnetic field
or to the problem of 
compound
states in an
atomic nucleus.  
As is usual for interacting many-body problems, 
evaluation of exact 
ground state is expensive task. 
So far, the only
results
for the yrast states
in repulsive case
were obtained numerically and for simplified 
$\delta$-forces\cite{BP}.

In this work, we provide 
rigorous 
analytical solution 
for the ground state
as a function of two quantum numbers,
the total angular momentum, 
$L$$($$\leq$$N$$)$,
and
the angular momentum of internal excitations,
for a broad class of repulsive interactions.
These {\it generalized yrast states} include the
usual yrast states. The results allow transparent
physical interpretation.
In fact,
the form of the yrast wave functions which
was drawn from numerics for the $\delta$-interaction case\cite{BP},
turns out to be valid in general.
The universality class of interactions
is described by an explicit sufficiency condition.

For this ``single Landau level'' problem,
the Hamiltonian 
($\hbar=m=1$) 
is the sum
\begin{eqnarray}\label{ham}
\tilde{H}= 
\omega H_0
+ \tilde{V},
\end{eqnarray}
where the first term 
(the energy of degenerate level)
comes from the noninteracting Hamiltonian 
in the parabolic trap,
$\sum_i^{N}
\left(\frac{\vec{p}_i^2}{2}+\frac{\omega^2}{2}\vec{r}_i^2 \right)$,
and the second (nontrivial) term $\tilde{V}$ is a properly projected
interaction $V=\sum\limits_{i>j} V(r_{ij})$,
see below.
In 2d, one uses complex $z(z^*)=x\pm iy$
instead of vector $\vec{r}=(x,y)$,
and we set $\omega\equiv 1$ for brevity.
With 
$\partial_{\pm}\equiv\frac{1}{2}(\frac{ \partial}{ \partial x}
\mp i\frac{ \partial}{ \partial y})$, 
we employ the tetrad of 
ladder
operators
$a^{+}$, $a$, $b^{+}$ and  $b$ for each
particle, 
\begin{eqnarray} \label{a-akrest}
a^{+} =  z/2-\partial_+, \quad 
b^{+} =  z^*/2-\partial_-, \quad 
\nonumber\\
a = (a^{+})^{\dagger}    , \quad
b = (b^{+})^{\dagger}     , 
\quad[a,a^{+}] = [b,b^{+}] = 1,
\end{eqnarray}
which raise (lower) the powers of $z_i$ and $z^*_j$ in the 
preexponentials of many-body wave functions, 
which are all polynomials times the Gaussian factor  
$|0\rangle=exp(-1/2\sum|z_k|^2)$.
So,
$z_i|0\rangle\equiv a^+_i|0\rangle$,
$z^*_j|0\rangle\equiv b^+_j|0\rangle$, $a_i|0\rangle=0$, 
$b_i|0\rangle=0$, {\it etc}. 
The 
2d
angular momentum is 
the difference $L=L_+ -L_-$ of numbers 
of ``up'' and ``down'' quanta,
$L_+=\sum_1^N a^{+}_ka_k$ 
and 
$L_-=\sum_1^N b^{+}_kb_k$,
while the energy of the level in (\ref{ham}) is $H_0=L_+ +L_-+N$. 
The yrast states must have minimum energy at given $L$.
Thus, they belong to the subspace
with $L_{-}=0$, $L_+=L$,
spanned 
by the homogeneous (degree $L$) symmetric polynomials ,
$
poly^{L}_S(a^{+}_i) |0\rangle
$
which do not involve $z^*$.
The basis is formed by
polynomials $[l_1,l_2,...,l_N]$,
obtained by 
applying the operator of symmetrization over $N$ variables, $P_S$,
to the monomials $m$ 
\cite{symmetricFUNCTIONS}
\begin{equation} \label{PARTITION}
[{l_1},{l_2},...,{l_N}] \equiv 
P_S
m, \qquad
m \equiv 
z_1^{l_1}z_2^{l_2}...z_N^{l_N}. 
\end{equation}
with 
a given partition of the integer 
$L=\sum\limits_n l_n$, 
\cite{symmetricFUNCTIONS}.
In our case, $L$$\leq$$N$, 
the dimensionality of 
the basis is
$p(L)$,
``the number of
unrestricted
partitions'' 
of integer $L$
\cite{ABR}.

The system enjoys an additional conserved
quantity\cite{Wilkin,Mottelson}.
The collective operators 
${\cal A}^{+}=\sum_{i=1}^{N}\frac{a^{+}_i}{\sqrt{N}}$
and 
${\cal A}\equiv({\cal A}^{+})^{\dagger}$ 
(with $[{\cal A}, {\cal A}^{+}]=1$)
commute 
with 
$\tilde{H}$, and so does the number 
of collective quanta, ${\cal A}^{+}{\cal A}$. 
The mutual eigenfunctions of $\tilde{H}$, $L$ and 
${\cal A}^{+}{\cal A}=v$ 
are factorized
\begin{eqnarray} \label{seniorityBASIS}
\Psi(L,v) = Z^{v} poly^{L-v}_S
(\tilde{z}_i)
e^{-\sum\frac{|z_k|^2}{2}},
\end{eqnarray}
with $\tilde{z}$$_i$$=$$z$$_i$$-$$Z$ 
and $Z$$\equiv$$\sum_1^N$$z_k$$/$$N$.
In (\ref{seniorityBASIS}), the
total angular momentum 
$L$
is redistributed 
between internal excitations ($J$$=$$L$$-$$v$) and
the collective motion ($v$).
It is therefore interesting to consider the ground 
state as a function of 
both
$L$ and $v$.
We call the eigenvalue of ${\cal A}^{+}{\cal A}$ 
{\it seniority} for brevity.
At given $L$, the allowed values 
are
$v=0,1,2,...,L-2,L$;
the value
$v=L-1$ is forbidden 
because $poly^{1}_S(z_i-Z)\equiv 0$.
The states with definite seniority, 
of type (\ref{seniorityBASIS}), can be obtained 
by applying the projector ${\cal P}_v$ \cite{PROJECTORS} 
to (\ref{PARTITION}).

In a given sector $L$, 
the projected\cite{PROJECTORS} interaction $\tilde{V}$
in (\ref{ham})
is given by\cite{HV1HV2,HV3}
\begin{eqnarray}\label{projection}
\tilde{V} = P_{L_-=0} P_{L_+=L} V P_{L_+=L} P_{L_-=0}. 
\end{eqnarray}
In order to evaluate (\ref{projection}), 
we use the Fourier representation
\begin{eqnarray}\label{fourier} 
V = 
\sum\limits_{i>j}
\int\limits_{-\infty}^{\infty}
d^2\vec{q}
e^{ i\left[ q_- 
(a^{+}_{ij}+b_{ij})+ q_+ (b^{+}_{ij}+a_{ij})
\right]} 
V_q 
\end{eqnarray}
where 
$V_q= 
\int_{0}^{\infty} rdr
J_0(r|q|)
V(r)/(2\pi)$ 
is the 2d Fourier transform of the potential $V(r)$, 
with $J_0$ the Bessel function \cite{ABR} and
$q$$_{\pm}$$=$$($$q$$_x$$\pm$$i$$q$$_y$$)$$/$$\sqrt{2}$.
The two-particle combinations 
$a^{+}_{ij}$$\equiv$$\frac{1}{\sqrt{2}}$$($$a^{+}_{i}$$-$$a^{+}_{j}$$)$,
$b^{+}_{ij}$$\equiv$$\frac{1}{\sqrt{2}}$$($$b^{+}_{i}$$-$$b^{+}_{j}$$)$
{\it etc.} came from resolving $x$$_{i(j)}$ and $y$$_{i(j)}$ from
(\ref{a-akrest}).
Substituting (\ref{fourier}) in (\ref{projection}) and using 
Baker-Hausdorf relation 
$e^{a^{+} - a}
= e^{\frac{[a^{+},a]}{2}} e^{a^{+}}e^{-a}$, 
we obtain 
expansion of $\tilde{V}$
in terms of the two-particle operators 
$B^k\equiv \sum\limits_{j>i}^{N} a^{\dagger k}_{ij} a_{ij}^k$, 
\begin{eqnarray}  \label{OPE}
\tilde{V} = \sum\limits_{k=0}^{2[L/2]}(-1)^ks_kB^{k} =
\sum\limits_{k=0}^{[L/2]} V_k,
\nonumber\\
V_k\equiv s_{2k}(B^{2k}-B^{2k-1})+
(s_{2k+2}-s_{2k+1})B^{2k+1}.
\end{eqnarray}
The highest possible order of $B^k$ is $L$ for $L$ even,
and $L-1$ for $L$ odd;
$B^0$$\equiv$$N$$(N$$-$$1$$)$$/$$2$.
The shape of 
potential $V(r)$ is described by 
the integrals with Kummer function $M$\cite{ABR},\cite{LAGUERRE}
\begin{eqnarray}  \label{OPE2}
s_k \equiv  
\int_0^{\infty}\frac{dt}{k!} M(k+1,1,-t) V(\sqrt{2t}).
\end{eqnarray}
Expansion (\ref{OPE}) is exact for any interaction
$V(r)$ whose moments $s_k$ are finite\cite{HV1HV2,HV3}.

Regular methods to obtain
exact ground state 
without solving the whole spectrum are not available.
We use the approach 
\cite{HV3} which we loosely nicknamed ``supersymmetry''
\cite{SUSYprimer}.
Suppose that the Hamiltonian 
can be written as a sum, 
\begin{eqnarray}\label{SUSY1}
\tilde{V}= V_0 + V_S,
\end{eqnarray}
such that 
\\
{\bf (a)} the first term, $V_0$, 
is simple, and one can
find out its ground state $|0)$
with eigenvalue ${\cal E}_{min}$ (possibly degenerate). 
If the second term, $V_S$,  
has
the two ``supersymmetric'' properties:

{\bf (b)}
$V_S$ annihilates the state $|0)$, 
so $V_S|0)=0$,

{\bf (c)} 
$V_S$ is {\it non-negative definite}, $V_S\geq 0$, (it does not have
negative eigenvalues),
\\
then 
the state $|0)$ will still be the ground 
state 
for the full Hamiltonian $V_0+V_S$, with 
the same eigenvalue ${\cal E}_{min}$.
Indeed, 
{\bf (b)} implies that $|0)$ is
an eigenstate for the sum $V_0+V_S$ with its eigenvalue
intact. 
As one knows form linear algebra, 
if an Hermitean 
operator 
$V_0$ is perturbed
by {\it a non-negative definite} Hermitean operator $V_S$, the 
eigenvalues can only increase (see, e.g. \cite{BOOK-MATRICES}).
This means that the states other than $|0)$
can gain energy [Cf. {\bf (b)}].
As $|0)$ is already the ground state for $V_0$, it
will be the same for $V_0+V_S$
\cite{exampleOSCILLATOR,COMMENTdegeneracy}.
 
The same arguments apply to the case with
additional conserved quantum number $v$, such as
$[$$v$$,$$V$$_0$$]$$=$$[$$v$$,$$V$$_S$$]$$=$$0$.
In the above scheme, the single state $|$$0$$)$ 
is replaced by the set of
states $|$$L$$,$$v$$)$,
each having the property {\bf (a)} and {\bf (b)} in their $v$-sectors.
This is illustracted in Fig.1, where the spectrum 
of $V_0$ is taken degenerate in each $v$-sector.

In our case, the supersymmetric triad $V_0$, $V_S$ and $|0)$
can be established by inspecting action of terms $V_k$
in (\ref{OPE}) on the states of partition basis (\ref{PARTITION}).
While the first term $V_0$, by virtue of identity
$N\sum_i^Na^+_ia_i=\frac{1}{2}\sum_{i,j}^N(a^+_i-a^+_j)(a_i-a_j)+
(\sum_i^Na^+_i)(\sum_i^Na_j)$,
is reduced to a combinations of quantum numbers,
\begin{eqnarray}\label{V0}
V_0=(N/2)[(N-1)s_0 -(L-{\cal A}^{+}{\cal A})(s_1 -s_2)],
\end{eqnarray}
the simplest state $|L\rangle$ in (\ref{PARTITION})
with partition $[1,1,...1]$ 
is annihilated by the remainder of the Hamiltonian,
\begin{eqnarray}\label{simplest}
|L\rangle = P_S a^+_1 a^+_2 ... a^+_L |0\rangle,
\qquad
(\tilde{V} - V_0) |L\rangle=0,
\end{eqnarray}
Indeed,  
$[a_{12}^2, a^+_1 a^+_2]=-2(a^{+}_{12}a_{12}+1)$
and thence $a_{12}^{k(>2)} a^+_1 a^+_2a^+_3...a^+_L|0\rangle=0$ 
and $B^{k>2}|L\rangle=0$.
From the same commutator, we have
$[a^{+2}_{12}a_{12}^2, a^+_1 a^+_2]|0\rangle=
[ a^{+}_{12}a_{12}, a^+_1
a^+_2]|0\rangle=-a^{+2}_{12}|0\rangle$,
and thus $(B^{2}-B^{1})|L\rangle=0$. 
Substitution 
$z_i=\tilde{z}_i+Z$ [see (\ref{seniorityBASIS})]
transforms $|L\rangle$ to a sum
\begin{eqnarray}\label{multivortex2}
|L\rangle = 
\biggl[
P_S \tilde{z}_1 \tilde{z}_2 ...\tilde{z}_L
+...
+Z^{L-2}P_S\tilde{z}_1 \tilde{z}_2
+ Z^{L} 
\biggr]|0\rangle =\sum\limits_v{\cal P}_v|L\rangle
\end{eqnarray}
of exactly $L$ {\it seniority observing states} of the form 
(\ref{seniorityBASIS}), 
each being the eigenvector 
of ${\cal A}^+{\cal A}$ and therefore of $V_0$, 
with the eigenvalue (\ref{V0}).
We notice that each $v$-sector is represented
by a single term in (\ref{multivortex2}),
identified with ${\cal P}_v|L\rangle$.

The spectrum of $V_0$ (see Fig.1.) consists of $L$ 
equidistant (except $v$$\neq$$L$$-$$1$), $g(v)$-fold 
degenerate levels with 
energies 
(\ref{V0});
$g$$($$v$$)$$=$$p$$($$L$$-$$v$$)$$-$$p$$($$L$$-$$v$$-$$1)$ 
for $v$$\leq$$L$$-$$2$ and $g$$($$L$$)$$=$$1$.
Each 
$v$-level contains one and only one state
$|L,v)={\cal P}_v|L\rangle$ from the sum (\ref{multivortex2})
Therefore, the set of 
states 
$|L,v)$
obey the criterion
{\bf (a)} with the operator $V_0$.
The property {\bf (b)} with $V$$_S$$\equiv$$\tilde{V}$$-$$V$$_0$
holds by virtue of (\ref{simplest}).
In particular, {\bf (a)} together with {\bf (b)} mean that 
$|L,v)$
are the eigenvectors of $\tilde{V}$$=$$V$$_0$$+$$V$$_S$ 
with eigenvalues (\ref{V0}).
This holds for any interaction $V(r)$.

The supersymmetric representation for the triad $V_0$,
$V_S$
and $|L,v)={\cal P}_v|L\rangle$
would be complete if we succeeded to prove
non-negative definiteness $V_S$$\geq$$0$ of 
the remainder of the Hamiltonian $\tilde{V}-V_0$,
criterion {\bf (c)}.
So far, we did not specify form of the interaction $V(r)$ in 
(\ref{OPE}).
We will study now general case and specify the 
class of potentials which have $V_S\geq 0$.

We have to check signs of all the eigenvalues of $V_S$
in the partition space (\ref{PARTITION}). 
To avoid solving the whole spectrum, 
in the space of symmetrized states,
we use the following
trick. 
By definition, the nonzero eigenvalues of 
$V_S$
in the space (\ref{PARTITION}) coinside
with the nonzero eigenvalues of $P_S V_S P_S$ in the 
{\bf full space of monomials} $m$ in (\ref{PARTITION}).
This latter space has dimensionality 
much higher than $p(L)$
and it includes wave functions of all possible symmetries,
including boson sector (fully symmetric), 
fermion sector (fully antisymmetric) {\it etc}.
In this extended space, the 
analysis of signs of eigenvalues 
is however crucially simplified, 
while the contributions from the symmetric sector can
be accurately separated.
Using the symmetry of $V_S$ under permutations of particles, 
we write
\begin{eqnarray}\label{INERTIA}
P_S V_S P_S =  
P_S \sum\limits_{i>j}V_{S,ij}  P_S 
= 
(N/2)(N-1)
P_S 
V_{S,12}P_S 
\end{eqnarray}
where $V_{S,ij}$ is the contribution from pair of particles $i$$,j$
to $V_S$ [Cf.(\ref{OPE},\ref{OPE2},\ref{SUSY1})].
In order to see that 
$P_S$$V_S$$P_S$$\geq$$0$, 
it is {\it sufficient}
to show that 
$V_{S,12}$$\geq$$0$,
because application of any 
projector 
$P_S$ 
from both sides in (\ref{INERTIA})
can add new zero eigenvalues, but can not add
negative eigenvalues.
This follows from the known ``inertia theorem'' of linear algebra
\cite{BOOK-MATRICES}.
Now, we study the eigenvalues of $V_{S,12}$.
Let $\pi_{12}$ be the operator of permutation of 
variables $1$ and $2$.
The triad 
\begin{displaymath}
T=\{\pi_{12}, \quad a^{+}_{12}   a_{12}, \quad V_{S,12} \}
\end{displaymath}
forms a set of mutually commuting operators. 
Indeed, $V_{S,12}$ 
is expressed 
in terms of $B^k_{12}=a^{+k}_{12}$$a^k_{12}$ 
[see (\ref{OPE})]. As
$[$$a_{12}$$,$$a^+_{12}$$]$$=$$1$,
any  $B^k_{12}$
can be expressed in terms of $a^+$$_{12}$$a_{12}$,
using boson calculus formula
$a^{+ k}_{12}$$a^k_{12}$$=
$$a^{+}_{12}$$a_{12}$$($$a^{+}_{12}$$a_{12}$$-$$1$$)$$...$
$($$a^{+}_{12}$$a_{12}$$-$$k$$+$$1$$)$. 
The triad $T$ is diagonalized simultaneously right in the 
basis of monomials
$m\equiv\{ z_1^{l_1} z_2^{l_2} z_3^{l_3}... z_N^{l_N} \}$,
(\ref{PARTITION})
with the only substitutions 
$z_1$$\rightarrow$
$\frac{1}{\sqrt{2}}$$($$z_1$$-$$z_2$$)$,
$z_2$$\rightarrow$$\frac{1}{\sqrt{2}}$$($$z$$_1$$+$$z$$_2$$)$.
In this basis, the eigenvalues of the triad $T$
depend only on $l_1$ in the subfactor
$\biggl[$$\frac{1}{\sqrt{2}}$$($$z$$_1$$-$$z$$_2$$)$$\biggr]$$^{l_1}$
of monomial $m$, they are
$\{$$(-1)^{l_1}$$,\quad$$l_1,$$\quad$$\lambda_{l_1}\}$, respectively.
Using Eqs.(\ref{OPE}) and (\ref{OPE2})
and the
summation formula
$\sum_{k=0}^N
\frac{(-1)^kN!}{k!(N-k)!}M(k+1,1,-t)=\frac{e^{-t}t^N}{N!}$,
we obtain 
$\lambda_{l_1}=
\int\limits_0^{\infty} r dr V(r)f_{l_1}(r)$ with 
$f_{l_1}(r)=e^{-r^2/2}
\left[ \frac{r^{2l_1}}{2^{l_1}l_1!} - 1 + \frac{l_1}{4}\left(2 -
\frac{r^4}{4}\right) \right]$.
The eigenvalue 
$($$-$$1$$)$$^{l_1}$ of $\pi$$_{12}$
helps now to separate out states with wrong symmetry:
the eigenvectors with $l_1$ odd are antisymmetric in $z_1,z_2$,
and the projector $P_S$
in (\ref{INERTIA}) kills them all. 
All even values of $l_1$ ($\leq$$L$) 
can contribute to the bosonic sector, and the corresponding
$\lambda$'s must be checked.

Now,
inequality $\lambda$$_{2n}$$\geq$$0$ 
for any $2$$n$$\leq$$L$
is the sufficient condition which defines the class
of potentials for which 
$V_S$ is 
non-negative definite, and the triad 
$V_0$, $V_S\equiv\tilde{V}-V_0$ 
and $\{ {\cal P}_v|L\rangle\}$
obeys the supersymmetric representation
with properties {\bf (a)},  {\bf (b)} and  {\bf (c)},
with ${\cal P}_v|L\rangle$ being the 
ground state
in its sector $L,v$ with the energy ${\cal E}_{min}$
equal to eigenvalue of $V_0$.

With this criterion and (\ref{multivortex2}) and (\ref{V0})
at hand, we can formulate very general result:
For any 
{\it bona fide} two-body
potential $V(r)$ which satisfies the integral condition
\begin{eqnarray}\label{RESULT1}
\int\limits_0^{\infty} V(\sqrt{2t}) e^{-t}
\left[ \frac{t^{2n}}{(2 n)!} - 1 + n\left(1 -
\frac{t^2}{2}\right) \right] \geq 0  
\nonumber\\
\qquad for \quad 
any \quad n\leq L'/2,
\end{eqnarray}
the eigenstates of the Hamiltonian 
$\tilde{H}$
{\bf with minimal
energies at 
given pair of $v$ and $L(\leq\min\{L',N\})$}
have universal form
\begin{eqnarray}\label{RESULT2}
|L, v ) = e^{-\frac{1}{2}\sum|z_i|^2} Z^v 
\left( \frac{\partial}{\partial Z}\right)^{N-L+v}
\prod\limits_{k=1}^{N} (z_k - Z), 
\qquad
\nonumber\\
\qquad
Z\rightarrow\frac{1}{N}\sum\limits_{i=1}^N z_i,
\end{eqnarray}
with the energies given by the simple moments of 
$V(r)$,
\begin{eqnarray}\label{RESULT3}
{\cal E}_{min}(L,v)= L + N + \frac{N(N-1)}{2}{\cal V}_0
+ \frac{ {\cal V}_0 - {\cal V}_1 }{2} N (v-L)
\nonumber\\
{\cal V}_0=\int_{0}^{\infty}dt e^{-t}V(\sqrt{2t}), \quad
{\cal V}_1=\int_{0}^{\infty}dt e^{-t}
\left(\frac{1}{2}+\frac{t^2}{4}\right)V(\sqrt{2t})
\end{eqnarray}
which
are equal to  
expectation values of interaction between
two bosons both in the ground 
${\cal V}$$_0$$=$$\langle$$0$$0$$|$$V$$|$$0$$0$$\rangle$ 
and the first excited state
of oscillator 
${\cal V}$$_1$$=$$\langle$$1$$1$$|$$V$$|$$1$$1$$\rangle$, 
respectively\cite{NEznachit}.

At fixed $L$, we have exactly $L$
such 
equidistant {\it generalized yrast}
states, marked by $v=0,1,2,...,L$ ($v\neq L-1$),
see Fig.1.
Each such state is the ``ground state'' in the sector $L,v$
(of course, there are other
states in each sector with higher energies).

The usual yrast states minimize ${\cal E}_{min}(L,v)$ 
with respect to seniority, $v$.
From (\ref{RESULT3}), they can have either $v=0$ or $v=L$.
As can be seen, in the domain of validity 
(\ref{RESULT1})
the first option is usually realized.
Inequality $({\cal V}_0 - {\cal V}_1)(v-L)\leq 0$ means
that internal rotational excitations are energetically 
favorable, once the interaction energy
between two bosons
in the state $z|0\rangle$ is less, then in the state $|0\rangle$.
Physically, the {\it yrast} wave functions (\ref{RESULT2}) with $v=0$
correspond to condensation to a vortex, rotating around the
``center-of-mass'', as discussed in \cite{Wilkin}.
The maximum seniority states $v=L$,
which correspond to purely collective rotation\cite{Mottelson}
with no internal excitations, were shown\cite{Wilkin} to
be energetically favorable for attractive $\delta$-forces.

Consider repulsive ($U_0$$\geq$$0$) Gaussian interaction, 
$V(r)=U_0 \frac{ e^{-r^2/R^2} }{\pi R^2}$,
whose range $R$ can be varied from zero to infinity.
From 
(\ref{RESULT1}), 
we have
$\lambda_{2n} = \frac{U_0}{\pi(2+R^2)}
[(\frac{R^2}{2+R^2})^{2n}-1+4n\frac{1+R^2}{(2+R^2)^2}]\geq0$,
so (\ref{RESULT1}) holds for any $R$, and the spectrum is
${\cal E}_{min}=L+N+\frac{U_0}{\pi(2+R^2)}
[N(N-1)/2 -  \frac{(1+R^2)}{(2+R^2)^2}N(L-v)  ]$. 
The yrast states have $v=0$.
In particular, in the zero range limit, $R$$\rightarrow$$0$,
we have the $\delta$-function repulsive interaction $V$$=$$U$$_0$
$\delta$$($$\vec{r}$$)$.  
We have
$\lambda_{l_1} = \frac{U_0}{2\pi}
[\delta_{l_1, 0} + \frac{l_1}{2}-1]$.
It is seen that 
while 
$\lambda$$_1$$=$$-$$1$$/$$2$, for any $l$$_1$$=$$2$$n$ even 
the condition 
$\lambda_{2n}\geq0$ (\ref{RESULT1}) holds. 
The energies
(\ref{RESULT3}) 
are ${\cal E}_{min}=L+N+\frac{U_0}{8\pi}N(2N-L+v-2)$.
Yrast states have $v=0$,
and ${\cal E}_{min}(L,0)$
agrees
with that obtained numerically \cite{BP}, see also \cite{dop}.
  
For the 2d Coulomb interaction $V=U_0log(1/r)$ with $U_0\geq 0$, 
we have
$\lambda_{2n}=\frac{U_0}{4}[3n-1/n+2\psi(1)-2\psi(2n)]\geq0$,
where $\psi$ is digamma function,
so 
(\ref{RESULT1}) holds, and
${\cal E}_{min}=L+N-\frac{U_0 N}{4}[ (log(2)-\gamma)(N-1) 
-(3/4)(v - L) ]$ with $\gamma=-\psi(1)=0.57721...$ the Euler constant.
The yrast states correspond to $v=0$.

Let us see, that condition (\ref{RESULT1}) imposes only weak
restrictions
on the repulsive forces $V$$($$r$$)$.
Indeed, at small $r\simeq 0$, 
the factor-function 
$f_{2n}$$($$r$$)$ in 
$\int_0^{\infty}$$r$$d$$r$$f_{2n}$$($$r$$)$$U$$($$r$$)$
in (\ref{RESULT1})
approach positive constant values,
$n-1$ and $df_n/dr|_{r=0}=0$, 
while the first node of the functions $f_{2n}(r)$ 
occurs at 
$r_0 = \sqrt{2\sqrt{12-2\sqrt{30} } }\simeq1.43$
which is of order of the oscillator length,
in our units ($\hbar$$=$$m$$=$$\omega$$=$$1$).
Therefore, for any short range 
(as compared to the characteristic length of the trap)
interaction, the condition (\ref{RESULT1}) 
reduces to 
\begin{displaymath}
\int d^2 \vec{r} V(r) \geq 0,  
\end{displaymath}
Thus, the results (\ref{RESULT2},\ref{RESULT3})
hold for
short-range interactions, which 
are repulsive on average. 

The condition (\ref{RESULT1}) holds even 
for many long-range interactions.
This is seen from behavior of function $f_4(r)$
($n=0$ and $n=1$ give $f\equiv0$)
which is positive at $r<r_0$ and 
$r>r_1=\sqrt{2\sqrt{12+2\sqrt{30} } }\simeq 3.10$,
and $f_4$ is negative only in the interval $r_0 \leq r \leq r_1$
($f_n(r)$ for higher $n$ behave similarly).
It is clear that (\ref{RESULT1}) holds, if $V(r)$ 
decreases monotonically and fast enough,
as is the case for 
the long range Gaussian
and Coulomb forces.
One can therefore summarize that
condition holds for any physically meaningful repulsive 
interaction.

To conclude, in the problem of weakly interacting
Bose condensate in 2d trap,
we developed an exact expansion for the 
interaction to powers of ladder operators.
A supersymmetric
representation for this interaction 
has been developed to obtain
the states with {\it minimum energy at given angular momentum
and seniority} (generalized yrast
states, including usual yrast states). 
The energies
of these condensed vortex states are given by 
simple integral moments
of the 
interaction potential. 
A broad {\it universality} class of the repulsive 
interactions for which these results are valid is described explicitly.

Further applications of these results are straightforward.
In view of\cite{3d2d}, the three-dimensional case
can be done using the same method.
It is also interesting to study region of higher angular momenta
$L>N$, where the structure of the basis polynomials will
be changed \cite{symmetricFUNCTIONS}, while the numerical studies
indicate signs of phase transition\cite{BP}.
The method of ``supersymmetric decomposition'' developed here is not
restricted to this particular problem and can be applied to 
fermions and even to the particles with parastatistics.
The work was supported by FAPESP.

\newpage
{\large Figure Captions}

Fig.1.
Illustraction of supersymmetric decomposition.
The spectrum of $V_0$ (left) is sequence of 
degenerate levels, labeled by the conserved
quantum number $v$. 
The spectrum of $V_0+V_S$
is shown on the right.
Supersymmetric perturbation $V_S$$\geq$$0$ splits each level,
pushing the states up and leaving the lowest energy
in each 
$v$-sector  intact. 


\begin{thebibliography}{300}
\bibitem{3d2d} As was shown in \cite{Mottelson,BP}, 
the 3d case reduces to 2d case in the weak coupling limit.
\bibitem{LAGUERRE}
For $k$ integer,
it
can be related to Laguerre polynomials 
$M(k+1,1,-t)=e^{-t}L^0_n(t)$.
\bibitem{exampleOSCILLATOR}
A trivial example is the 1d harmonic oscillator 
with $V=1/2+a^+a$. The state
$|0)$, such as $a|0)=0$, is ``the ground state'' for
$V_0\equiv 1/2$ because any other state $|n)$ has the same
eigenvalue $1/2$. The operator $V_S=a^+a$, $V_S\geq0$ acts
as perturbation, removing
this degeneracy and leaving $|0)$ with lowest eigenvalue of
$V$\cite{COMMENTdegeneracy}.
\bibitem{COMMENTdegeneracy} In principle, this does not exclude 
possibility that
the ground state can still be degenerate, 
if it was such
for $V_0$. 
It is possible to show\cite{HV3} that
in our case 
the ground states of $\tilde{V}$ are not degenerate.
\bibitem{NEznachit}
This does not mean that (\ref{RESULT3}) can be obtained as
an expectation value over a simple state
with only two oscillator levels occupied; in fact,
the eigenstates (\ref{RESULT2}) 
are much more complicated in the second quantization representation.
\bibitem{dop}
When the paper was in preparation, we have learned on recent works 
\cite{nov} where the states (\ref{RESULT2}) with $v$$=$$0$ were
shown to belong to the eigenspace 
for the case of $\delta$-interaction.
Generalization on the case $v$$\neq$$0$ was considered
in the last work of Ref.\cite{nov}.
\end{thebibliography}
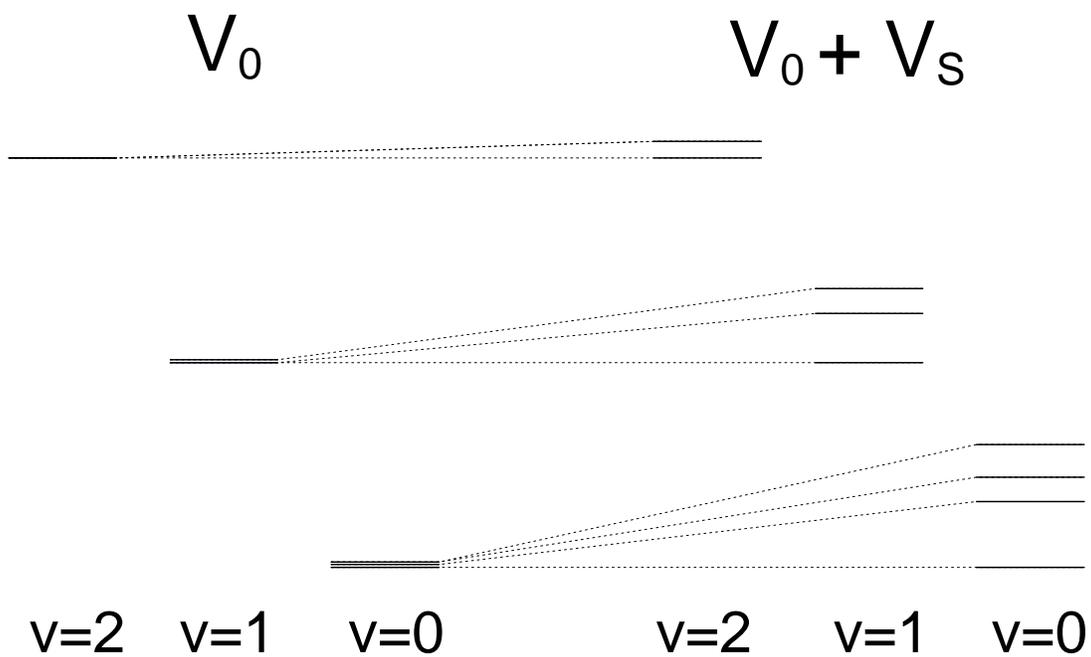
\end{document}